\documentclass[12pt]{article}

\usepackage{amsmath,amssymb}

\newcommand{\be}{\begin{eqnarray}}
\newcommand{\ee}{\end{eqnarray}}
\newcommand{\nn}{\nonumber}
\newcommand{\m}{\mu}
\newcommand{\n}{\nu}
\renewcommand{\r}{\rho}
\newcommand{\s}{\sigma}

\newcommand{\p}{\partial}
\renewcommand{\o}{\omega}
\newcommand{\eps}{\epsilon}
\newcommand{\tA}{\tilde{A}}

\begin{document}

\rightline{UG-08-03}
\rightline{ULB-TH/08-08}
\rightline{KCL-MTH-08-03}

\begin{center}
{\huge \sc Dual Gravity and Matter}
\end{center}
\vskip8mm
\begin{center}
Eric A. Bergshoeff\footnotemark[1], 
Mees de Roo\footnotemark[1], 
Sven F. Kerstan\footnotemark[1]${}^{,}{}$\footnote[2]{Present affiliation: 
Optiver, De Ruyterkade 112, 1011 AB Amsterdam, The Netherlands},\\ 
 Axel Kleinschmidt\footnotemark[3] and Fabio Riccioni\footnotemark[4]\\[6mm] 
\footnotemark[1]{\sl  \small Centre for Theoretical Physics, University of
  Groningen,\\ 
  Nijenborgh 4, 9747 AG Groningen, The Netherlands}\\[3mm]
\footnotemark[3]{\sl \small Physique Th\'eorique et Math\'ematique \&{}
  International Solvay Institutes ,\\
Universit\'e Libre de Bruxelles,\\
Boulevard du Triomphe, ULB -- CP 231, B-1050 Bruxelles, Belgium}\\[3mm]
\footnotemark[4]{\sl \small  Department of Mathematics, King's College
  London,\\ 
  Strand, London WC2R 2LS, United Kingdom} \\[7mm]

\begin{tabular}{p{11cm}}
\hspace{5mm}{\footnotesize {\bf Abstract:} 
We consider the problem of finding a dual formulation of gravity in the
presence of non-trivial matter couplings. In the absence of matter a dual
graviton can be introduced only for linearised gravitational interactions. We
show that the coupling of linearised gravity to matter poses obstructions to the usual construction and comment on possible resolutions of this difficulty.}

\end{tabular}\\[5mm]

\end{center}

\begin{section}{Introduction\label{Intro}}

One of the remarkable features of $D=4$ electrodynamics is that it allows for
both an electric formulation, using the vector potential $A_\mu$, and for a
magnetic formulation, using a dual vector potential $\tilde{A}_\mu$, in any
background described by a metric $g_{\mu\nu}$. The duality relation between
these two fields can be written as 
\be 
F_{\mu\nu} = \frac12 \eps_{\mu\nu\rho\tau} \tilde{F}^{\rho\tau}\,,
  \quad F_{\mu\nu} = 2\partial_{[\mu}A_{\nu]}\,,
  \quad \tilde{F}_{\mu\nu} = 2\partial_{[\mu}\tilde{A}_{\nu]}\,.
\label{vectorduality}
\ee
The integrability condition for the local existence of $\tilde{A}_\mu$ is the
original field equation: $-\frac12
\eps^{\nu\s_1\s_2\s_3}\nabla_{\s_1}\tilde{F}_{\s_2\s_3} = \nabla_\mu
F^{\mu\nu} = 0$. Generally, the duality exchanges field equations and Bianchi
identities.  The duality property can be preserved when the Maxwell
field is coupled to other matter, e.g., axion/dilaton scalar fields,
but breaks down when generalised to non-abelian gauge groups.
The construction (\ref{vectorduality}) in $D=4$ generalises to any $p$-form $A_p$ (that is any field  
with $p$ antisymmetric spacetime indices) in $D$ dimensions, the dual  
of the field $A_p$ being a $(D-p-2)$-form $\tilde{A}_{D-p-2}$.

It is natural to ask whether a similar dual formulation exists for the
gravitational field. For {\em linearised} gravity in vacuo
\footnote{For a 
discussion
of gravitational duality in (Anti) de Sitter space see 
\cite{Julia:2005ze, Leigh:2007wf}.} such a
formulation is known to exist
\cite{Curtright:1980yk,Nieto:1999pn,West:2001as,Hull:2001iu,
  West:2002jj,Bekaert:2002dt,
  Henneaux:2004jw,Ajith:2004ia,Nurmagambetov:2006vz,Ellwanger:2006hy} but a BRST analysis
reveals,
under rather general assumptions \cite{Bekaert:2002uh}, 
obstructions to extend this to a theory with covariant and local 
interactions.  

Expanded around a flat background the metric takes the form
\be
g_{\mu\nu} = \eta_{\mu\nu} + \kappa h_{\mu\nu} + O(h^2)
\ee
and the curvature tensors simplify in linear order to
\be
R_{\mu\nu\,\rho\sigma} = 2\p_{[\mu} \omega_{\nu]\,\rho\sigma}\,,
   \quad R_{\mu\nu} = - \p_\mu \omega_{\rho\,\rho\nu} - \p_{\rho}
   \omega_{\mu\,\nu\rho}\,, 
   \quad R= -2 \p_\rho \omega_{\sigma\,\sigma\rho}\,,
\ee
where now all derivatives are partial and indices are raised and lowered with
the flat Minkowski metric and we disregard higher order terms in the graviton
$h_{\mu\nu}$ from now on.  Evidently, the curvature tensors are of
order $O(\kappa h)$.\footnote{The dimensions are:
  $[\kappa]=\frac{2-D}{2}$, $[h_{\mu\nu}]=\frac{D-2}{2}$,
  $[\omega_{\mu\,\nu\rho}]=1$ and $[R_{\mu\nu\,\rho\sigma}]=2$.}
 The spin connection is $\omega_{\mu\,\nu\rho} = 2\kappa
\p_{[\nu}h_{\rho]\mu}$ in terms of the graviton and satisfies
$\omega_{[\mu\,\nu\rho]}=0$.  The linearised vacuum Einstein equations in
$D$-dimensional space-time can be written as \cite{West:2001as}
\be\label{einvac}
0 = R^\nu{}_\rho -\tfrac12\delta^\nu_\rho R = - \frac{1}{(D-2)!}
\eps^{\nu\s_1\ldots \s_{D-1}} 
  \p_{\s_1} Y_{\s_2\ldots \s_{D-1},\rho} \,,
\ee
where
\be\label{dualvac}
Y_{\mu_1\ldots \mu_{D-2}, \rho} = \tfrac12\eps_{\mu_1\ldots
  \mu_{D-2}}{}^{\s_1\s_2}  
  \left( \omega_{\rho\,\s_1\s_2} -2 \eta_{\rho\s_1}
    \omega^\nu{}_{\nu\s_2}\right)  
\ee
is obtained from dualising the spin connection and its trace. $Y$ is contained
in the tensor product of a vector with a $(D-2)$-form, we use a comma to
seperate the antisymmetric indices from the single vector index. Equation
(\ref{einvac}) suggests the introduction of a dual graviton
$D_{\mu_1\ldots \mu_{D-3}, \rho}$ via
\be
\label{dualgraviton}
(D-2)\p_{[\mu_1} D_{\mu_2\ldots \mu_{D-2}],\rho} = Y_{\mu_1\ldots \mu_{D-2},
  \rho} 
\ee
as solution to the $Y$-Bianchi identity (\ref{einvac}), which is equivalent to the graviton
equation of motion. The consequence of linearisation
$\omega_{[\mu\, \nu\rho]}=0$ is equivalent to $Y_{\mu_1\ldots
  \mu_{D-3}\nu}{}^\nu=0$, which is a differential condition on the dual
graviton. It was argued in \cite{West:2002jj} that the 
condition $D_{[\mu_1\ldots\mu_{D-3},\rho]}=0$ can be imposed by a local Lorentz transformation.
The equations (\ref{einvac}) and (\ref{dualvac})  can be derived from the 
Einstein action in
first order formulation as shown in \cite{West:2001as}: One introduces $Y$ as
an auxiliary field in the action which then depends on the vielbein and
$Y$. Substituting the solution of the algebraic equation 
of motion for $Y$ gives back the Einstein action. 
In this framework (\ref{dualvac}) is the algebraic equation of motion for $Y$ whereas (\ref{einvac}) is the
equation of motion obtained by varying with respect to the vielbein and
linearising (see also \cite{West:2002jj}).   

A slightly different approach for the introduction of a dual graviton starts
from the Riemann tensor and its symmetries
\cite{Curtright:1980yk,Hull:2001iu,Bekaert:2002dt}. Dualising the full linearised Riemann
tensor $R_{\m\n\r\s}$ on one set of antisymmetric indices gives the tensor  
\be
 S_{\m_1\ldots \m_{D-2}\,\r\s} = \tfrac12 \eps_{\m_1\ldots
    \m_{D-2}}{}^{\n_1\n_2} R_{\n_1\n_2\,\r\s} \,.
\ee 
The (algebraic and differential) identities for the Riemann tensor together
with the linearised equations of motion then imply that on-shell \cite{Hull:2001iu,Bekaert:2002dt}
\be\label{sint}
S_{\m_1\ldots \m_{D-2}\,\r\s} &=& \p_\s\p_{[\m_1} \tilde{D}_{\m_2\ldots
    \m_{D-2}],\r} -\p_\r\p_{[\m_1} \tilde{D}_{\m_2\ldots
      \m_{D-2}],\s}\nn\\
&=& \p_\s \tilde{Y}_{\m_1\ldots \m_{D-2},\r} - \p_\r
    \tilde{Y}_{\m_1\ldots \m_{D-2},\s} 
\ee
in terms of a dual graviton $\tilde{D}_{\m_1\ldots \m_{D-3},\r}$
which manifestly satisfies $\tilde{D}_{[\m_1\ldots
    \m_{D-3},\r]}=0$. The linearised Einstein equation in this case
is obtained by taking antisymmetric parts of $S$, e.g.
\be\label{einrievac}
\frac{1}{(D-3)!}\eps^{\mu\s_1\ldots \s_{D-1}} 
   S_{\nu\s_1\ldots \s_{D-1}} 
  = R^\mu{}_\nu - \tfrac12\delta^\mu_\nu R \,.
\ee
In this approach
there is no local duality relation similar to
(\ref{dualvac}). Arguably the best one can hope for is  
\be\label{dualrie}
\tilde{Y}_{\m_1\ldots \m_{D-2},\r} =\tfrac12\eps_{\m_1\ldots
  \m_{D-2}}{}^{\s_1\s_2} \o_{\r\,\s_1\s_2} + \p_\r
\tilde{\Lambda}_{\m_1\ldots \m_{D-2}},
\ee
where $\tilde{\Lambda}_{\m_1\ldots \m_{D-2}}$ is a possibly non-local
term which ensures that all symmetry properties are satisfied. The term
$\tilde{\Lambda}_{\m_1\ldots \m_{D-2}}$ is allowed for since it drops out in
$S$, cf.~(\ref{sint}).

This paper is organised as follows. In Section \ref{SUSY} we will show
that the dual graviton can also be introduced in the context of linearized
supergravity in $D=4$. Our approach uses the duality relation (\ref{dualrie}).
In Section \ref{Matter} we discuss dual gravity in  the presence of 
gravity and matter, in an arbitrary number of dimensions, and determine 
the conditions on the energy-momentum tensor that this matter coupling 
requires. The analysis of these
conditions shows that linearised gravity and dual gravity cannot be combined 
with matter. In Section \ref{Disc} we discuss these results and possible
escape routes.

\end{section}

\begin{section}{Supersymmetry in $D=4$\label{SUSY}}

In this section we show that the supersymmetry algebra of minimal supergravity in four dimensions closes on
the dual graviton $\tilde{D}_{\mu\nu}$ at the linearised level.
At lowest order in the fermions, the supersymmetry
transformations of the vielbein and the gravitino are
\begin{eqnarray}\label{sugrasusy}
  \delta e_\mu{}^a &=& \tfrac{1}{2}\,\bar\epsilon\,\gamma^a\psi_\mu\,,
\nonumber\\
  \delta \psi_\mu &=&
    (\partial_\mu - \tfrac{1}{4}\omega_\mu{}_{\alpha\beta}\gamma^{\alpha\beta})\epsilon
    \,,
\end{eqnarray}
where the spinor $\epsilon$ and the gravitino are Majorana. We want to linearise gravity around a flat
background, and this corresponds to considering linearised global supersymmetry transformations
\begin{eqnarray}
  \delta h_{\mu\nu} &=& \bar\epsilon\gamma_{(\mu}\psi_{\nu)}\,,
\nonumber\\
  \delta \psi_\mu   &=& -\tfrac{1}{4} \gamma^{\alpha\beta}\omega_{\mu\,\alpha\beta}\epsilon
         \,,
\label{linsusy}
\end{eqnarray}
where $\omega_{\mu\, \alpha \beta}$ is the linearised spin connection, and $h_{\mu\nu}$ is the first order
fluctuation of the metric.

In four dimensions the dual graviton has the same spacetime index structure as the graviton, and thus we
denote it with $\tilde{D}_{\mu\nu}$, where the spacetime indices are meant to be symmetrised. This field varies with
respect to general coordinate transformations, that at the linearised level are translations, but it also
possesses its own gauge transformations, that have the form
  \begin{equation}
  \delta \tilde{D}_{\mu\nu} = \partial_{(\mu} \Lambda_{\nu )} \,,\label{dualgravgaugetransfD=4}
  \end{equation}
where $\Lambda_\mu$ is an arbitrary gauge parameter. This gauge transformation has precisely the same
structure as the general coordinate transformation of the linearised graviton. This would not be true in
dimensions other than four.

We require the supersymmetry transformation of the dual 
graviton to be
  \begin{equation}
  \delta \tilde{D}_{\mu\nu} = \frac{i}{2} \bar\epsilon \gamma_{(\mu} \gamma_5 \psi_{\nu )} \,,
  \end{equation}
where in our conventions 
$\gamma_5 = -i \gamma_0 \gamma_1 \gamma_2 \gamma_3$, and 
we are using mostly +
signature. Using eq. (\ref{linsusy}), the commutator of 
two supersymmetry transformations on $\tilde{D}_{\mu\nu}$
gives
  \begin{equation}
  [\delta_1 , \delta_2 ] \tilde{D}_{\mu\nu}= - \frac{i}{4} \omega_{(\mu}{}^{\alpha\beta} \bar\epsilon_2 \gamma_{ \nu) \alpha\beta} \gamma_5 \epsilon_1
  = - \frac{1}{2}  \omega_{(\mu }{}^{\alpha\beta} \epsilon_{ \nu) \alpha\beta \gamma} \xi^\gamma
 \, ,\label{dualgravcommutator}
  \end{equation}
where
  \begin{equation}
  \xi^\mu = \frac{1}{2} \bar\epsilon_2 \gamma^\mu \epsilon_2
  \end{equation}
is the general coordinate transformation parameter that occurs in the commutator of two supersymmetry
transformations on the graviton and on the gravitino.

In this four dimensional case, the duality relation (\ref{dualrie}) becomes
  \begin{equation}
  \tilde{Y}_{\mu\nu, \rho} + \partial_\rho \tilde\Lambda_{\mu\nu} = \frac{1}{2} \epsilon_{\mu\nu\alpha\beta}
  \omega_{\rho}{}^{\alpha\beta} \,,
  \end{equation}
where
  \begin{equation}
  \tilde{Y}_{\mu\nu , \rho} = \partial_\mu \tilde{D}_{\nu\rho} - \partial_{\nu} \tilde{D}_{\mu\rho} \, .
  \end{equation}
Using these equations, eq. (\ref{dualgravcommutator}) becomes
  \begin{equation}
  [\delta_1 , \delta_2 ] \tilde{D}_{\mu\nu} = \xi^\gamma \partial_\gamma \tilde{D}_{\mu\nu} - \xi^\gamma \partial_{(\mu}
  \tilde{D}_{\nu)\gamma} - \xi^\gamma \partial_{(\mu} \tilde\Lambda_{\nu)\gamma} \, .
  \end{equation}
Given that at the linearised level we can treat $\xi$ as a constant, this result shows that this
supersymmetry commutator produces a gauge transformation as in eq. (\ref{dualgravgaugetransfD=4}), with
parameter
  \begin{equation}
  \Lambda_\mu= -\xi^\gamma (\tilde{D}_{\mu\gamma} + \tilde\Lambda_{\mu\gamma} )\,,
  \end{equation}
as well as translations. This proves that one can close the supersymmetry algebra of minimal supergravity in
four dimensions on the dual graviton at the linearised level.
\end{section}

\begin{section}{Inclusion of matter\label{Matter}}

Matter couples to gravity via its energy-momentum tensor
\be
R_{\mu\nu}-\frac12 g_{\mu\nu} R = \kappa^2 T_{\mu\nu} \,.
\ee
One can retain non-linear matter while linearising gravity. At
lowest non-vanishing order in the graviton, matter and gravity decouple and
one is left with the sum of a free spin two field and the remaining, possibly
self-interacting, matter propagating on a Minkowski background. In this
situation one can dualise the graviton as before since there are no matter
contributions in the defining equations. This trivial dualisation is, however,
not satisfactory from the point of view of the recently proposed
infinite-dimensional symmetries \cite{West:2001as} where the dual graviton
should bear some marks of the matter present in the theory.\footnote{Indeed, 
in the example of $D=11$ supergravity one would expect from the structure of 
the $E_{11}$ coset element that the dual graviton transforms non-trivially 
under the gauge transformations of the three-form potential and its dual 
six-form and that these transformations cannot be completely removed by 
field redefinitions.}

Repeating the steps that led to (\ref{dualgraviton}) in the matter coupled action yields 
again the duality relation (\ref{dualvac}), but now (\ref{einvac}) is
replaced by
\be
\label{einmat}
\p_{[\m_1}Y_{\m_2\ldots \m_{D-1}],\r} &=& 
   \kappa^2 \tilde{T}_{\m_1\ldots \m_{D-1},\r}\,, 
\ee
where the right hand side in (\ref{einmat}) is dual to the energy
momentum tensor $T_{\mu\nu}$:
\be\label{dualem}
\tilde{T}_{\m_1\ldots \m_{D-1},\r} =\frac{(-1)^{D-2}}{(D-2)!}
   \eps_{\m_1\ldots \m_{D-1}}{}^\s T_{\s\r} \, .
\ee
The symmetry of $T_{\mu\nu}$ implies that the trace of the dual
energy-momentum tensor vanishes. Now, since the r.h.s. of (\ref{einmat}) is no
longer zero we are not immediately led to the introduction of a dual 
graviton $D_{\m_1\ldots \m_{D-3},\r}$; the integrability condition
has changed. If, however, the dual of the energy-momentum tensor satisfies
\be\label{emcurl}
\tilde{T}_{\m_1\ldots \m_{D-1},\r} = -\p_{[\m_1}M_{\m_2\ldots \m_{D-1}],\r}\,,
\ee
which is equivalent to
\be
\label{emcurl1}
  T^\lambda{}_\r = \frac{(-1)^{D-2}}{D-1} \epsilon^{\m_1\ldots\m_{D-1}\lambda}
      \p_{\m_1}M_{\m_2\ldots \m_{D-1},\r}\,,         
\ee
we can define an improved $Y$ by
\be
\label{Yimprove}
Y_{\m_1\ldots \m_{D-2},\r} \rightarrow Y_{\m_1\ldots \m_{D-2},\r} 
   + M_{\m_1\ldots \m_{D-2},\r} \,.
\ee 
This improved $Y$ then satisfies the standard integrability relation and gives 
rise to the dual graviton as before. This improvement is only useful if $M$
has a local expression in the matter fields and their duals. In other words,
the introduction of a dual graviton in the presence of matter is equivalent to
peeling one derivative off the dual energy momentum tensor in
(\ref{emcurl}).  

A similar conclusion is reached by studying the approach via the Riemann
tensor. To obtain the Einstein equation as an integrability condition from
(\ref{einrievac}) one requires that $S$ gives rise to the energy-momentum
contribution from the matter sector. This requires that 
there exists a tensor $\tilde{M}$ which plays the same role with respect to 
$\tilde{Y}$ as $M$ to $Y$ in (\ref{Yimprove}):
\be
\label{einmat2}
\tilde{Y}_{\m_1\ldots \m_{D-2}\,\r} \to 
   \tilde{Y}_{\m_1\ldots \m_{D-2}\,\r} + \tilde{M}_{\m_1\ldots\m_{D-2}\,\r}\,,
\ee
which again leads to the problem of finding a local expression
$\tilde{M}$ such that the Einstein equation arises from (\ref{einrievac}). 

We have investigated, in a variety of cases related to supergravity
systems with hidden symmetries, the relation (\ref{emcurl}) for 
the dual energy-momentum tensor to obtain 
local expressions for $M$ and $\tilde{M}$. 
For simplicity we present the analysis in
$D=4$ with gravity coupled to a single Maxwell field $A_\mu$ with the covariant
energy-momentum tensor  
\be
T_{\mu\nu} = F_{\mu\s_1} F_{\nu}{}^{\s_1} - \tfrac14 g_{\mu\nu} F_{\s_1\s_2}
F^{\s_1\s_2} \,. 
\ee
In lowest order the dual energy-momentum tensor (\ref{dualem}) then takes the
form
\be\label{dualmax}
\tilde{T}_{\mu_1\mu_2\mu_3,\rho} 
   = \tfrac34 F_{\rho[\mu_1} \tilde{F}_{\mu_2\mu_3]} 
     -\tfrac34 \tilde{F}_{\rho[\mu_1} F_{\mu_2\mu_3]} 
     + \tfrac32 \eta_{\rho[\mu_1} F_{\mu_2}{}^\s \tilde{F}_{\mu_3]\s} \,.
\ee
Since $T^\mu{}_\mu=0$ here we also have the constraint that
$\tilde{T}_{[\mu_1\mu_2\mu_3,\rho]}=0$. According to (\ref{emcurl}) we make
the ansatz
\be\label{Man}
M_{\mu_1\mu_2,\rho} &=& \alpha_1 A_{[\mu_1} \p_{\mu_2]}\tA_{\rho} 
   + \alpha_2 A_{[\mu_1} \p_{|\rho|}\tA_{\mu_2]}
   + \alpha_3 A_{\rho} \p_{[\mu_1}\tA_{\mu_2]}\nn\\
&& + \beta_1 \tA_{[\mu_1} \p_{\mu_2]}A_{\rho} 
   + \beta_2 \tA_{[\mu_1} \p_{|\rho|}A_{\mu_2]}
   + \beta_3 \tA_{\rho} \p_{[\mu_1}A_{\mu_2]} \\
&& + \gamma_1 \eta_{\rho[\mu_1} A_{\mu_2]}\p^\nu \tA_\nu 
   + \gamma_2 \eta_{\rho[\mu_1} A^\n\p_{\mu_2]} \tA_\nu
   + \gamma_3 \eta_{\rho[\mu_1} A^\n\p_{|\nu|}  \tA_{\mu_2]}   \nn\\
&&+ \gamma_4 \eta_{\rho[\mu_1} \tA_{\mu_2]}\p^\nu A_\nu 
   + \gamma_5 \eta_{\rho[\mu_1} \tA^\n\p_{\mu_2]} A_\nu
   + \gamma_6 \eta_{\rho[\mu_1} \tA^\n\p_{|\nu|}  A_{\mu_2]}   \,,\nn
\ee
without any restrictions on the real coefficients $\alpha_i$, $\beta_i$ $\gamma_i$.\footnote{Demanding that $M_{\mu_1\mu_2,\rho}$ comes from the dual graviton requires that $M_{[\mu_1\mu_2,\rho]}=0$, or $\alpha_1-\alpha_2+\alpha_3=\beta_1-\beta_2+\beta_3=0$ but we relax this condition for the moment.}
The terms with coefficients $\alpha_i$ and $\beta_i$ are needed to reproduce the first two terms in (\ref{dualmax}) whereas the $\gamma_i$ terms in the ansatz correspond to the third term in (\ref{dualmax}). Taking a curl of (\ref{Man}) through $\p_{[\mu_1} M_{\mu_2\mu_3],\rho}$ and
demanding that all terms combine into covariant field strengths after
dualisation implies for $\alpha_i$ and $\beta_i$ that
\be
\alpha_1+\beta_3 = 0 \,,\quad \alpha_3+\beta_1 =0 \,,\quad
\alpha_2 =0 \,,\quad \beta_2 =0 \,
\ee
and all $\gamma_i=0$. Any $M_{\mu_1\mu_2,\rho}$ satisfying this condition leads to
$\p_{[\mu_1} M_{\mu_2\mu_3],\rho}=0$ which implies $\tilde{T}_{\mu_1\mu_2\mu_3,\rho}=0$. Therefore one cannot recover the matter
coupled Einstein equations from a dual formulation in this
way.\footnote{Allowing for a term which is a total $\rho$ derivative as in
  (\ref{dualrie}) there are additional possibilities and there is a solution
  which gives the first two terms in (\ref{dualmax}). The third term cannot be
  accounted for in this way.}

Turning to the introduction of the dual graviton via the dualised Riemann
tensor as in (\ref{sint}) one can again use the ansatz (\ref{Man}) for
$\tilde{M}_{\mu_1\mu_2,\rho}$. Now the matter coupled
Einstein equation should arise as in (\ref{einrievac}), which leads to
the following condition between $\tilde{M}$ and the energy-momentum tensor:
\be
\label{emcurl2}
\tfrac12 \eps^{\mu\s_1\s_2\s_3} \p_{\s_3} \tilde{M}_{\nu\s_1,\s_2} 
  = T^\mu{}_\nu \,.
\ee
Without making any assumptions on the symmetry of $\tilde{M}_{\mu_1\mu_2,\rho}$
one finds a one-parameter family of non-trivial solutions represented by
\be
\alpha_1 =-\alpha_3 = \tfrac1{15} \,,\quad \alpha_2 = 1 \,,\quad
\beta_1 =-\beta_3 = \tfrac13 \,,\quad  \beta_2 =\tfrac15 \,.
\ee
All coefficients can be rescaled by the same constant.
However, insisting on the irreducibility condition of the dual graviton
(which automatically holds in the approach through the dualised Riemann
tensor), removes this solution. This difficulty was already anticipated in
\cite{Bekaert:2002dt}. 

The result of the explicit analysis above can be summarized in the following way 
\cite{Nurmagambetov:2008qs}. If we could find a solution for $M$ in (\ref{emcurl1})
(or $\tilde{M}$ in (\ref{emcurl2}))
the energy-momentum tensor would be defined in terms of
a local improvement term, and would be conserved independently of the equations
of motion. This is clearly undesirable.

\end{section}

\begin{section}{Discussion\label{Disc}}

In both approaches to the dual graviton we found that there is no satisfactory
way of coupling linearised gravity to matter and then describing both the
gravity and the matter sector using dual variables in a local and covariant
way. This is reminiscent of the findings of \cite{Deser:1969wk,Bunster:2006rt}
where it was also argued that the coupling of linearised gravity to dynamical
matter sources induces a non-linear completion of the gravity sector. Treating
the gravity sector non-linearly, one is however immediately faced with the
problem of the obstructions established in \cite{Bekaert:2002uh} when trying
to maintain locality and covariance. One possible way out then is to
abandon covariance \cite{Ellwanger:2006hy}, see also \cite{Damour:2002cu}.

One of the motivations for this work was to add the dual graviton to the
supersymmetry algebra in eleven dimensions in the same spirit as was done 
for the dual matter fields in $D=10$ maximal supergravity theories in
\cite{Bergshoeff:2005ac,Bergshoeff:2006qw}. The supersymmetry algebra 
in $D=10$ closes
on the dual matter fields if one imposes appropriate duality equations which
imply the dynamical matter equations of supergravity. This computation can
also be done using algebraic correspondences \cite{Damour:2006xu} and it is
therefore tempting to use the same techniques to derive the supersymmetry
rules of the dual graviton coupled to matter in maximal supergravity. 
If successful, this would reveal 
the way the dual graviton transforms under the $A_{(3)}$ and $A_{(6)}$ gauge 
transformations as required by supersymmetry, 
which could then be compared to the predictions of, e.g., $E_{11}$.
Whereas the dual graviton of pure minimal supergravity in $D=4$ 
can be included in the supersymmetry
algebra if one linearises and uses a duality relation of the type
(\ref{dualrie}) (see Section \ref{SUSY}) we find that in $D=11$ 
matter enters the duality relation in such a way that it no longer gives 
rise to the
correct, gauge invariant Einstein-matter equations. Phrased differently, the
supersymmetry algebra can be closed on the dual graviton in maximal
supergravity (and the answer agrees with the algebraic considerations) but
the duality relation is not an equivalent reformulation of the Einstein
equation. This result is in agreement with the
non-existence of a dual graviton coupled to matter using the approach we
outlined in Section \ref{Matter}. 

Finally we discuss some possible resolutions of this apparent
difficulty in addition to abandoning Lorentz covariance  which was already
mentioned. A possible but trivial resolution is to fully decouple the matter 
and the gravity 
sector (as suggested by a $\kappa$
expansion of the equations) and treat them as sums of free fields\footnote{This is
what happens also in Kaluza--Klein reduction. Linearised pure gravity in $D$
dimensions admits a dual graviton. After dimensional reduction to $D-1$
dimensions this gives again dual gravitation but the Kaluza--Klein scalar and
vector do not couple to gravity in $D-1$ dimensions.}. One should keep in mind that
there are (at least) two ways to introduce the dual graviton, as presented
in Section \ref{Intro}. Additional possibilities or combinations might be envisaged, and
it would be useful to understand the precise relation between these different
approaches.
The way the Einstein equations were constructed from the tentative
dual graviton involved very specific choices of taking derivatives,
cf.~(\ref{einvac}) and (\ref{einrievac}). Since the dual graviton is a mixed
symmetry tensor there be might other curvatures one could construct from it which
then give the Bianchi identities and field equations of the original
theory. However, this has to be done in such a way that the
assumptions of the generalised Poincar\'e lemma \cite{Bekaert:2002dt} are
satisfied, and we
could not find any non-trivial solution this way. This leads us to conclude 
that the requirement of a local and covariant expression for $M$ ($\tilde{M}$) 
cannot be maintained.

\end{section}

\vskip5mm
\noindent{\bf Acknowledgements}
\vskip5mm
The authors gratefully acknowledge discussions with  
Marc Henneaux, Olaf Hohm, Hermann
Nicolai and Peter West.
AK and FR would like to thank the University of Groningen for its 
hospitality during
several visits. AK is a Research Associate of the Fonds de 
la Recherche Scientifique--FNRS, Belgium.
The work of FR is supported by a PPARC rolling
grant PP/C507145/1 and the EU Marie Curie research 
training network grant MRTN-CT-2004-512194.

\end{document}